\documentstyle[12pt,aaspp4,epsf]{article}

\newcommand{\ltap}{\ \raisebox{-.4ex}{\rlap{$\sim$}} \raisebox{.4ex}{$<$}\ }
\newcommand{\gtap}{\ \raisebox{-.4ex}{\rlap{$\sim$}} \raisebox{.4ex}{$>$}\ }
\begin{document}
\slugcomment{\tt Preprint SISSA 129/96/EP/A}
\title{ Kicked Neutron Stars and Microlensing}
\author{Silvia Mollerach and Esteban Roulet}
\affil{Scuola Internazionale Superiore di Studi Avanzati \\
Via Beirut 4, I-34013 Trieste, Italy.}

\begin{abstract}

Due to the large kick velocities with which neutron stars are born in
supernovae explosions, their spatial distribution is more
extended than that of their progenitor stars. The large
scale height of the neutron stars above the disk plane makes
them potential candidates for microlensing of stars in the Large
Magellanic Cloud. Adopting for the distribution of kicks 
the measured velocities of young pulsars, we obtain a microlensing
optical depth of $\tau \sim 2 N_{10} \times 10^{-8}$ (where $N_{10}$
is the total  number of neutron stars born in the disk in units of
$10^{10}$). The event duration distribution has the interesting
property of being peaked at $T \sim 60~ $--$~ 80$ d, but for the rates
to be relevant for the present microlensing searches would require
$N_{10} \gtap 1$, a value larger than the usually adopted ones ($N_{10}
\sim 0.1~$--$~ 0.2$).
 
\end{abstract}

The ongoing searches of baryonic dark matter in the Galaxy by means of
microlensing of stars in the Large Magellanic Cloud (LMC) have
produced very surprising results recently. Indeed, the first 
five events obtained by the EROS (Aubourg {\it et al.} 1993) and MACHO
(Alcock {\it et al.} 1995) collaborations, with
durations $T \sim 10~ $--$~29 $ d, suggested the presence of compact objects
with masses  $\sim 0.1 M_\odot$, i.e. near the brown dwarf range. However,
the new analysis of the first two years of MACHO observations  (Alcock
{\it et al.} 1996) has now returned
events with longer durations (besides eliminating their two shortest
 duration events). These new results are indicative of
larger lens masses,
in the range $0.1~$--$~1 M_\odot$, being responsible for the events.
In addition, the large event rates observed require a considerable
amount of compact objects to be present in the Galaxy.

A first difficulty associated with the new range of inferred masses
is that the lensing population can no longer be completely ascribed to
the unseen brown dwarf counterpart of one of the Galactic stellar
components, such as the thick disk (Gould, Miralda--Escud\'e and
Bahcall 1994) or the spheroid (Giudice, Mollerach and Roulet 1994). 
Although a population of white dwarf remnants in these components 
could help to fit the observed MACHO event durations, it would require
an initial mass function quite contrived in order to have $\gtap 90
\%$ of the mass of these components ending up in white dwarfs, and
even in this case the resulting rates would still fall short with 
respect to the observed ones.
On the other hand, the proposed interpretation of the lenses being
an old white dwarf population in the dark halo makes the problem of the
absence of any luminous stellar counterpart more severe than for the
previous scenarios.

In this letter, we want to study the relevance for the microlensing
searches of the other natural candidate for dark 
lensing objects in the Galaxy: the neutron stars (NS).
An important point to take into account in the study of the population
of neutron stars in the Galaxy is the fact that the violent birth of
these objects in supernova (SN) explosions makes them acquire large `kick'
velocities, as evidenced by the observed velocities of young pulsars
(fastly rotating magnetized neutron stars). The pulsar velocities have
been studied by Lyne and Lorimer (1994), who obtained a fit to the
observed distribution of transverse speeds $v_t$ given by
\begin{equation}
\label{vel2d}
f_{2D}(v_t)\propto \frac{t^{0.13}}{1+t^{3.3}},
\end{equation}
where $t=v_t/(330$ km/s). Under the assumption that the kick
distribution is isotropic, the two dimensional distribution is just a
convolution of the underlying three dimensional distribution
$f_{3D}(v)$, 
\begin{equation}
f_{2D}(v_t)=v_t \int_{v_t}^\infty dv
\frac{f_{3D}(v)}{v\sqrt{v^2-v_t^2}}.
\end{equation}
Hence, $f_{3D}$ can be reconstructed by solving an Abel integral
equation, to obtain\footnote{The result obtained here is equivalent to 
the one of Terasawa and Hattori (1995).}
\begin{equation}
\label{vel3d}
f_{3D}(v)=-\frac{2}{\pi}\int_v^\infty dv_t \frac{df_{2D}}{dv_t}
\frac{v_t}{\sqrt{v_t^2-v^2}}. 
\end{equation}
The average velocity resulting from this distribution is $\langle v
\rangle \sim 390$ km/s. This
implies that the NS born from a given Galactic population
(e.g. the disk or the bulge) will end up with a completely different,
and much more extended, spatial distribution. The physical origin of
these kick velocities is still unknown, and proposed mechanisms include
the unbinding of a close binary system (Dewey and Cordes 1987, Bailes
1989), an asymmetric explosion (Burrows {\it et al.} 1995)
 or the interaction of the emitted
neutrinos with the NS magnetic field (Kusenko and Segr\`e 1996).

The spatial distribution of the NS born from massive disk
stars during the whole lifetime of the Galaxy has been considerably
studied as a possible Galactic source of gamma ray bursts (Shklovskii
and Mitrofanov 1985, Li and
Dermer 1992, Podsiadlowski, Rees and Ruderman 1995). The neutron
stars density can be obtained from Monte Carlo simulations of the
initial distribution of neutron star locations and kick velocities at
birth, and then following their orbits in the Galactic gravitational
potential up to the present (Paczy\'nski 1990, Hartmann, Epstein and
Woosley 1990, Terasawa and Hattori 1995). In particular, in order to 
fit the isotropy of the BATSE observations (Meegan {\it et al.} 1992), 
the gamma ray bursts in these Galactic scenarios must  
be produced by the very distant tail of the spatial NS
distribution, i.e. at Galactocentric distances $r \gtap 100$ kpc.
As we will show, a large fraction of the neutron stars born in the
disk ($\sim 30 \%$) contributes to the density at radii 8 kpc $< r < 40$
kpc (having a considerable height above the disk plane). Hence,
independently of their possible relevance to account for the gamma
ray bursts, the
NS can effectively act as lensing objects in the microlensing searches
in the direction of the LMC.

To compute the NS distribution we follow closely the work
of Paczy\'nski (1990), but with the currently accepted velocity
distribution, eqs. (\ref{vel2d},\ref{vel3d}). We take for the radial
distribution of the initial birth positions
\begin{equation}
dP (R) \propto \exp(-R/h_R) R dR,
\end{equation}
taking for the disk scale length $h_R=3.5$ kpc. For simplicity, we
take z$=0$ for the initial NS vertical positions, since
anyhow very massive disk stars form with a small scale height ($<
100$ pc), so that this assumption is not essential for the microlensing
results. With these distributions, we produced $1.6 \times 10^5$ 
initial values of positions and velocities and followed the orbits by 
integrating the equations of motion (using cylindrical coordinates)
\begin{eqnarray}
\frac{dR}{dt}&=v_R, \ \ \ \ \ \ \ \ 
&\frac{dv_R}{dt}=-\frac{\partial\Phi}{\partial R}+\frac{J_z^2}{R^3},\\
\frac{d{\rm z}}{dt}&=v_z, \ \ \ \ \ \ \ \ 
&\frac{dv_z}{dt}=-\frac{\partial\Phi}{\partial {\rm z}}
\end{eqnarray}
in the Galactic gravitational potential $\Phi$. 
In addition, we have the equations stating the conservation of the
angular momentum and the energy
\begin{eqnarray}
J_z&=&R v_\phi= {\rm const},\nonumber\\
E&=&v_R^2+v_z^2+v_\phi^2+\Phi(R,{\rm z})= {\rm const}.
\end{eqnarray}
In particular, we used the energy conservation to check that the final
accuracy for this quantity after the numerical integration was better
that $10^{-6}$.

We model the Galaxy with three dynamical components, a disk, a bulge 
and a dark halo, with $\Phi=\Phi_d+\Phi_b+\Phi_h$. For the
disk and bulge components we used Miyamoto and Nagai (1975) 
potentials given by
\begin{equation}
\label{gpot}
\Phi_i=-\frac{G
M_i}{\sqrt{R^2+\left(a_i^2+\sqrt{{\rm z}^2+c_i^2}\right)^2}}, 
\ \ \ \ \ \ (i=d,b).
\end{equation}
For the halo we took a logarithmic potential (Binney and Tremaine
1987)
\begin{equation}
\Phi_h=\frac{1}{2} v_0^2
\ln\left(R_c^2+R^2+\frac{{\rm z}^2}{q_\phi^2}\right) + {\rm const},
\end{equation}
 allowing in principle for a non--spherical distribution ($q_\phi
\neq 1$) in order to explore the effects of a flattened halo. The
corresponding mass densities can be obtained from the Poisson equation
$\nabla^2 \Phi = 4\pi G \rho$. The parameters in eqs. (\ref{gpot}) are
chosen so as to produce an acceptable rotation curve, and are
$M_d=8.07 \times 10^{10} M_\odot$, $a_d=3.7~ {\rm kpc}$,
$c_d=0.2~ {\rm kpc}$
for the disk and
$M_b=1.12 \times 10^{10} M_\odot$, $a_b=0$, 
$c_b=0.227~ {\rm kpc}$
for the bulge, with $v_0=200$~km/s and $R_c=10$~kpc for the halo
potential.

The resulting local column density, for a solar Galactocentric
distance of 8.5~kpc, is $\Sigma_0(|$z$|< 1.1$ kpc$)= 79
M_\odot/$pc$^2$ (for $q_\phi =1$), which is consistent with the
dynamical estimates from vertical motion of stars (Kuijken and Gilmore
1991), and increases to $\sim 85 M_\odot/$pc$^2$ for $q_\phi = 0.8$ 
(which corresponds to and axis ratio in the
density distribution $q_\rho \simeq 0.4$ (Binney and Tremaine 1987)). 
The rotation curve velocity, $v_c^2 = R \partial\Phi/\partial R|_{z=0}$, is
clearly insensitive to $q_\phi$.

Regarding the total number of neutron stars produced during the whole
Galaxy lifetime, this is a quantity which is not easy to estimate, since
it depends crucially on the initial mass function of heavy stars ($m >
8 M_\odot$). This is quite uncertain, mainly because of the unknown
evolution of the star formation rate from the birth of the Galaxy up
to our days. If the present accepted rate of Type II and Ib (core
collapse) supernovae of $\sim 0.02 - 0.03$ yr$^{-1}$, which is
consistent with the estimated pulsar birth rate (Lyne, Manchester and
Taylor 1985), were to be representative
of the average SN rate in the past, one would estimate a total number
of neutron stars produced in the disk $N \simeq 3 \times
10^8$. However, the star formation rate was certainly higher at
earlier times, mainly due to the larger fraction of gas present
in the past and to the possibility of an initial burst of star
formation.  This can enhance a lot the fraction contributed by the 
heavy stars to the initial mass function (Larson 1986). 
For instance, a recent modeling of the SN production in the Galaxy 
(Timmes, Woosley and Weaver 1995, 1996) concluded that a more
reasonable value for the total number of neutron stars is $N_{NS}
\simeq 1.9 \times 10^9$, but the uncertainty in this number is
undoubtedly large. In view of this, we will not adopt a definite value
for $N_{NS}$, giving the results in terms of $N_{10}=N_{NS}/ 10^{10}$,
hoping that in the future this number will be known more precisely.
Since in the models leading to $N_{10} \gtap 0.1$ most of
the SN explosions took place in the early life of the Galaxy, we just
assumed for the Monte Carlo simulations that the NS were all born 
$10^{10}$ years ago. The results are in any case nearly unchanged for 
a uniform distribution of birth times.

Figure 1 shows the number density contours of the resulting NS 
distribution. This distribution is more extended than the one obtained
by Paczy\'nski (1990) and Hartmann, Epstein and Woosley (1990), due to
the smaller velocities adopted at the time by those authors. We have 
indicated also in the plot the coordinates of the line of sight to the
LMC (which position at $(R,$z$)_{LMC}=(42,-26.5)$ kpc is
also labeled), assuming a distance to it of $D_{os}=50$ kpc. Looking
at this curve, it is useful to recall that using the parameter $x
\equiv D_{ol}/D_{os}$ to describe the normalized distance along the
line of sight to an hypothetical lensing object, at distance $D_{ol}$
from the Sun, one has $x=$z/z$_{LMC}$.

From the resulting NS density we can now compute the
microlensing optical depth (for reviews on microlensing see Roulet and
Mollerach (1996) and Paczy\'nski (1996))
\begin{equation}
\tau= \int_0^1 dx \frac{d\tau}{dx}, 
\end{equation}
with 
\begin{equation}
\frac{d\tau}{dx}=\frac{4\pi G}{c^2}D_{os}^2 x (1-x) \rho (x),
\end{equation}
where $\rho$ is the mass density of the lenses, for which we just
assume a common NS mass of $1.4 M_\odot$. The resulting value is
$\tau = 1.9 N_{10} \times 10^{-8}$.
Also, defining
\begin{equation}
\langle D_{ol}\rangle =\frac{1}{\tau}\int_0^1 dx \frac{d\tau}{dx} D_{ol},
\end{equation}
we find an average distance to the lenses of $\langle D_{ol}\rangle
\simeq 8.8$ kpc, which shows that the lenses are indeed quite far away
(we recall that $\langle D_{ol}\rangle \simeq 1.1, 3.6$ and 14 kpc for
objects in a thin disk, thick disk and dark halo respectively).

Another important quantity to describe the microlensing events is the
distribution of event durations. To obtain it one needs the distribution
of the velocities with which the lensing objects cross the line of 
sight to the LMC. To this end, we obtained the velocities of the NS
which were near to the line of sight, computed the relative velocities
with respect to this line, and then took the projections of these
relative velocities orthogonal to the same line.
If ${\bf v}_o$ and ${\bf v}_s$ are the velocities of the Sun and
the LMC respectively, the motion of the line of sight (and hence of the
so-called microlensing tube) is just
${\bf v}_t = (1-x){\bf v}_o + x{\bf v}_s$.
In a coordinate system where $\hat{\bf x}$ points toward the Galactic
center and $\hat{\bf y}$ along the direction of increasing longitudes,
we adopted, following Griest (1991), ${\bf v}_o=(9,231,16)$ km/s and
${\bf v}_s=(53,-160,162)$ km/s. The quantity of interest is
\begin{equation}
v_\perp =|{\bf v}_r - ({\bf v}_r \cdot \hat{L})\hat{L}|,
\end{equation}
where ${\bf v}_r ={\bf v}_l - {\bf v}_t$ is the relative velocity,
 with ${\bf v}_l$ being the lens
velocity which we obtained with the Monte Carlo. $\hat{L}=(\cos b \cos
\ell, \cos b \sin \ell, \sin b )$ is the versor in the direction of
the LMC, which has Galactic coordinates $(b,\ell)_{LMC}=(-33^\circ,281^\circ)$.

The duration of the events is given by $T \equiv R_E/v_\perp$,
with the Einstein radius being $R_E=2 \sqrt{G m D_{os} x(1-x)}/c$. 
The inclusion of the tube velocity is relevant in this scenario, contrary to
the case of lenses in a halo, mainly due to the importance of the
observer's motion and the `memory' that the NS have of their
progenitor's motion of rotation.

We can compute now the `theoretical' rate of events (assuming unit
efficiency) from
\begin{equation}
\langle\Gamma\rangle_{th}=\int_0^1 dx
\frac{d\langle\Gamma\rangle_{th}}{dx}, 
\end{equation}
where
\begin{equation}
\frac{d\langle\Gamma\rangle_{th}}{dx}=\frac{2}{\pi} \frac{d\tau}{dx}
\frac{1}{\langle T \rangle (x)},
\end{equation}
with $\langle T\rangle (x) =R_E (x) \langle v_\perp^{-1} \rangle (x)$.
The resulting value is $\langle \Gamma \rangle_{th} = 0.4 N_{10}$
events$/(10^7$ stars yr). We can equally compute the `theoretical'
average event duration from the relation
\begin{equation}
\langle T\rangle={2\over \pi}{\tau\over \langle\Gamma\rangle_{th}},
\end{equation}
resulting in $\langle T\rangle\simeq 115$~d.

The event duration distribution can be
obtained from the underlying distribution of velocities $v_\perp$ as a
function of $x$. We have computed it by dividing the event durations
in bins of $\Delta T= 10$~d, i.e. considering the intervals
$[T_i,T_i+\Delta T]$ with $T_i=\Delta T
\times i$, and obtaining the fraction $f_i(x)$ of the events with
durations in the specified intervals to estimate
the differential distribution $df(x,T)/dT$ of this fraction.
The differential rate can then be computed using
\begin{equation}
\frac{d\Gamma}{dT}(T)=\frac{2}{\pi}\int_0^1 dx \frac{d\tau}{dx}
\frac{1}{T} \frac{df(x,T)}{dT}.
\end{equation}
The resulting distribution is shown in Figure 2. A distinctive feature
of it is that it peaks at $T \sim 70$ d, a fact that is
interesting in view of the long duration of the 
events recently
obtained by the MACHO collaboration (Alcock $et$ $al.$ 1996). 
However, one has also to recall that
for their accumulated statistics of $\sim 1.8 \times 10^7$ stars yr, and
adopting an average efficiency in this range of event durations
$\epsilon \sim 0.3$, one would expect a number of events $N_{ev}
\simeq  0.3 N_{10}$ arising from the NS.

We have also explored the sensitivity of the results to variations in
the assumptions made. Considering a flattened halo changes little
the previous predictions. This is because the effects of flattening
only affect sizeably the distribution of NS along the line of sight 
for $|$z$|>10$ kpc, and the
contribution to the optical depth of lenses in this region is not
dominant in
any case.

Variations in the radial distribution of initial positions can have
some effects on the predictions. For instance, adopting a Gaussian
distribution centered at 5 kpc with an extension of 2 kpc, as done
by Hartmann {\it et al.} (1990), increases the optical depth by 
$\sim 20 \%$. Considering
NS produced in the inner region of the Galaxy, as e.g. resulting from
SN explosions of stars in the Galactic bulge, leads to a smaller
optical depth `per star borned'\footnote{For instance, $\tau 
\simeq 0.8\times 10^{-8} N_{10}$ for NS born in the inner 1~kpc of the
Galaxy.}, with events having somewhat larger  durations.

There is little sensitivity of the predictions to the details of the
assumed velocity distribution, and
for instance halving all the initial velocities leaves almost
unchanged both $\tau$ and $\langle T\rangle$. Most of
the contribution to $\tau$ comes from NS with intermediate velocities
$100$ km/s $\ltap v \ltap 500$ km/s. For example, if all stars were
given initial kicks of 50, 100, 300 or 600 km/s, we would obtain an
optical depth $\tau= 0.5, 1.6, 3.0, 0.5 \times 10^{-8} N_{10}$
respectively.

Finally, we would like to note that another logical possibility is 
that a fraction of the SN
explosions leads to black hole (BH) formation rather than to NS.
For instance, Timmes {\it et al.} (1996)  have estimated that 
if all stars more massive than $19
M_\odot$ gave rise to BH rather than to NS, $\sim 70 \%$ of the
remnants would be BH ($N_{BH}\simeq 1.4 \times 10^9$ according to
their estimates). Two important differences would result in this
case. The kick velocities imparted to the BH in the explosions, if
any, would certainly be smaller, leading to a  BH distribution
less extended than the one obtained for the NS, and this would tend
to reduce the optical depth toward the LMC. Second, the larger average
masses of the produced BH would tend to enhance the optical depth
proportionally ($\tau \propto \rho \propto \langle m_{BH}\rangle$). If
a significant fraction of the BH acquire velocities larger than $\sim
100$ km/s, the first effect will not be large, as previously discussed,
so that the second one would dominate, allowing for an increased value
of $\tau$. The larger masses would also imply longer event
durations ($T \propto \sqrt{m}$).

In conclusion, we have considered the contribution that NS may provide
to the microlensing of LMC stars. Leaving aside the possibility of a
pregalactic (population III) NS component (Eichler and Silk 1992,
Brainerd 1992), 
the largest NS population
would be the one arising from SN explosions in the Galactic disk,
since this is the most massive stellar Galactic component. The
resulting optical depth for this population is $\tau \sim 2 \times 
10^{-8} N_{10}$, and the event durations are peaked at $T \sim 70$
d. This range of durations is interesting in
view of the long event durations observed by the
MACHO collaboration, with the longest two having $T=65$ and 
50~d\footnote{The determination of the duration is
sensitive to the treatment of the blending effects, and for these two
events the unblended fit gives $T=57$ and 43~d respectively. Also note
that there is no neat distinction between short and long duration
events.} (a larger one, with $T=71$ d and produced by a binary
lens\footnote{Due to the fact that the large majority of NS are
single, a characteristic of the NS events will be the lack of binarity 
signals}, most probably
corresponds to a lens in the LMC, due to its measured proper motion,
while the events of shorter durations may (at least partially) be 
ascribed to a faint component of some of the Galactic or LMC stellar 
populations (see de R\'ujula {\it et al.} 1995)).
However, for the reference value $N_{10}\simeq 0.2$, the expected
event rate from the NS falls below the observed one by an order of
magnitude (although the statistics  is certainly not large). 
Only for a larger NS density, $N_{10} \gtap 1$, the rates from NS
would be significant for the present microlensing searches. 
Whether such large values are acceptable and
consistent with the Galactic chemical evolution remains to be seen. 
At any rate, the existence of this lensing population certainly made
it deserve a quantitative study.
\bigskip\bigskip

We thank A. Lanza for useful discussions.


\newpage

\figcaption[figmns1.ps]{ Present neutron star density contours
vs. cylindrical galactic coordinates. The different contours
correspond to $N/N_{10}=$3.E-3, 1.E-3, 3.E-4, 1.E-4, 3.E-5, 1.E-5
(stars/pc$^3$). The location of the LMC is indicated as well as the
coordinates (solid line) of the line of sight to that
galaxy.\label{fig1}}

\figcaption[figmns2.ps]
{Differential event rate distribution vs. event
duration, taking as normalization $N_{10}=1$.\label{fig2}}

\clearpage

\begin{figure}
\center{\epsfbox{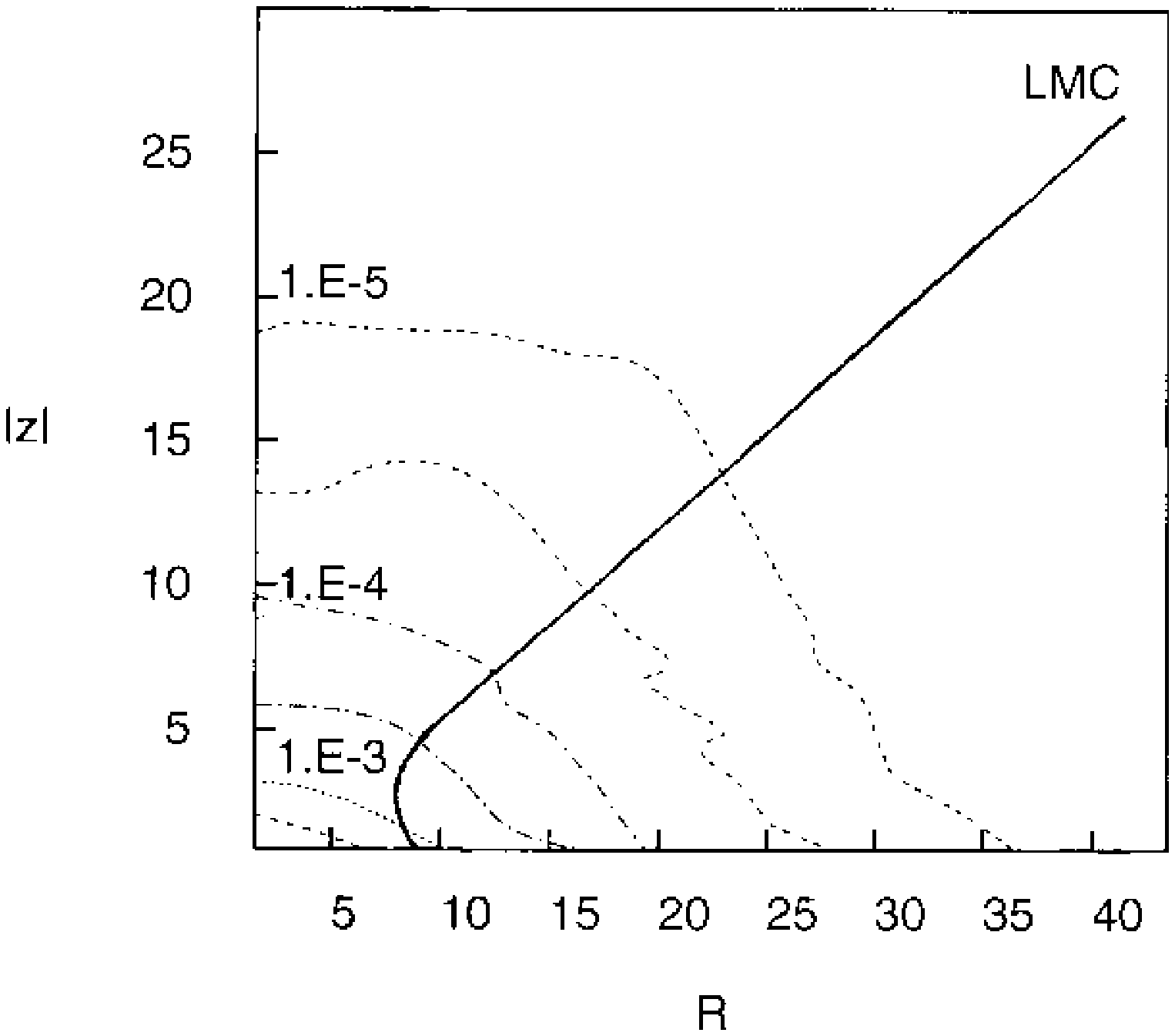}}
\end{figure}

\clearpage

\begin{figure}

\setlength{\unitlength}{0.240900pt}
\ifx\plotpoint\undefined\newsavebox{\plotpoint}\fi
\sbox{\plotpoint}{\rule[-0.200pt]{0.400pt}{0.400pt}}%
\begin{picture}(1500,900)(0,0)
\font\gnuplot=cmr10 at 10pt
\gnuplot
\sbox{\plotpoint}{\rule[-0.200pt]{0.400pt}{0.400pt}}%
\put(176.0,113.0){\rule[-0.200pt]{303.534pt}{0.400pt}}
\put(176.0,113.0){\rule[-0.200pt]{0.400pt}{173.207pt}}
\put(176.0,113.0){\rule[-0.200pt]{4.818pt}{0.400pt}}
\put(154,113){\makebox(0,0)[r]{0}}
\put(1416.0,113.0){\rule[-0.200pt]{4.818pt}{0.400pt}}
\put(176.0,216.0){\rule[-0.200pt]{4.818pt}{0.400pt}}
\put(154,216){\makebox(0,0)[r]{0.0005}}
\put(1416.0,216.0){\rule[-0.200pt]{4.818pt}{0.400pt}}
\put(176.0,318.0){\rule[-0.200pt]{4.818pt}{0.400pt}}
\put(154,318){\makebox(0,0)[r]{0.001}}
\put(1416.0,318.0){\rule[-0.200pt]{4.818pt}{0.400pt}}
\put(176.0,421.0){\rule[-0.200pt]{4.818pt}{0.400pt}}
\put(154,421){\makebox(0,0)[r]{0.0015}}
\put(1416.0,421.0){\rule[-0.200pt]{4.818pt}{0.400pt}}
\put(176.0,524.0){\rule[-0.200pt]{4.818pt}{0.400pt}}
\put(154,524){\makebox(0,0)[r]{0.002}}
\put(1416.0,524.0){\rule[-0.200pt]{4.818pt}{0.400pt}}
\put(176.0,627.0){\rule[-0.200pt]{4.818pt}{0.400pt}}
\put(154,627){\makebox(0,0)[r]{0.0025}}
\put(1416.0,627.0){\rule[-0.200pt]{4.818pt}{0.400pt}}
\put(176.0,729.0){\rule[-0.200pt]{4.818pt}{0.400pt}}
\put(154,729){\makebox(0,0)[r]{0.003}}
\put(1416.0,729.0){\rule[-0.200pt]{4.818pt}{0.400pt}}
\put(176.0,832.0){\rule[-0.200pt]{4.818pt}{0.400pt}}
\put(154,832){\makebox(0,0)[r]{0.0035}}
\put(1416.0,832.0){\rule[-0.200pt]{4.818pt}{0.400pt}}
\put(176.0,113.0){\rule[-0.200pt]{0.400pt}{4.818pt}}
\put(176,68){\makebox(0,0){0}}
\put(176.0,812.0){\rule[-0.200pt]{0.400pt}{4.818pt}}
\put(302.0,113.0){\rule[-0.200pt]{0.400pt}{4.818pt}}
\put(302,68){\makebox(0,0){20}}
\put(302.0,812.0){\rule[-0.200pt]{0.400pt}{4.818pt}}
\put(428.0,113.0){\rule[-0.200pt]{0.400pt}{4.818pt}}
\put(428,68){\makebox(0,0){40}}
\put(428.0,812.0){\rule[-0.200pt]{0.400pt}{4.818pt}}
\put(554.0,113.0){\rule[-0.200pt]{0.400pt}{4.818pt}}
\put(554,68){\makebox(0,0){60}}
\put(554.0,812.0){\rule[-0.200pt]{0.400pt}{4.818pt}}
\put(680.0,113.0){\rule[-0.200pt]{0.400pt}{4.818pt}}
\put(680,68){\makebox(0,0){80}}
\put(680.0,812.0){\rule[-0.200pt]{0.400pt}{4.818pt}}
\put(806.0,113.0){\rule[-0.200pt]{0.400pt}{4.818pt}}
\put(806,68){\makebox(0,0){100}}
\put(806.0,812.0){\rule[-0.200pt]{0.400pt}{4.818pt}}
\put(932.0,113.0){\rule[-0.200pt]{0.400pt}{4.818pt}}
\put(932,68){\makebox(0,0){120}}
\put(932.0,812.0){\rule[-0.200pt]{0.400pt}{4.818pt}}
\put(1058.0,113.0){\rule[-0.200pt]{0.400pt}{4.818pt}}
\put(1058,68){\makebox(0,0){140}}
\put(1058.0,812.0){\rule[-0.200pt]{0.400pt}{4.818pt}}
\put(1184.0,113.0){\rule[-0.200pt]{0.400pt}{4.818pt}}
\put(1184,68){\makebox(0,0){160}}
\put(1184.0,812.0){\rule[-0.200pt]{0.400pt}{4.818pt}}
\put(1310.0,113.0){\rule[-0.200pt]{0.400pt}{4.818pt}}
\put(1310,68){\makebox(0,0){180}}
\put(1310.0,812.0){\rule[-0.200pt]{0.400pt}{4.818pt}}
\put(1436.0,113.0){\rule[-0.200pt]{0.400pt}{4.818pt}}
\put(1436,68){\makebox(0,0){200}}
\put(1436.0,812.0){\rule[-0.200pt]{0.400pt}{4.818pt}}
\put(176.0,113.0){\rule[-0.200pt]{303.534pt}{0.400pt}}
\put(1436.0,113.0){\rule[-0.200pt]{0.400pt}{173.207pt}}
\put(176.0,832.0){\rule[-0.200pt]{303.534pt}{0.400pt}}
\put(806,23){\makebox(0,0){$T$ [d]}}
\put(366,877){\makebox(0,0){${d\Gamma/ dT}$ [events/(yr $10^7$ stars d)]}}
\put(176.0,113.0){\rule[-0.200pt]{0.400pt}{173.207pt}}
\put(176,113){\usebox{\plotpoint}}
\multiput(176.00,113.58)(2.538,0.495){35}{\rule{2.100pt}{0.119pt}}
\multiput(176.00,112.17)(90.641,19.000){2}{\rule{1.050pt}{0.400pt}}
\multiput(271.58,132.00)(0.499,0.835){123}{\rule{0.120pt}{0.767pt}}
\multiput(270.17,132.00)(63.000,103.409){2}{\rule{0.400pt}{0.383pt}}
\multiput(334.58,237.00)(0.499,1.289){123}{\rule{0.120pt}{1.129pt}}
\multiput(333.17,237.00)(63.000,159.658){2}{\rule{0.400pt}{0.564pt}}
\multiput(397.58,399.00)(0.499,1.289){123}{\rule{0.120pt}{1.129pt}}
\multiput(396.17,399.00)(63.000,159.658){2}{\rule{0.400pt}{0.564pt}}
\multiput(460.58,561.00)(0.499,1.074){123}{\rule{0.120pt}{0.957pt}}
\multiput(459.17,561.00)(63.000,133.013){2}{\rule{0.400pt}{0.479pt}}
\multiput(523.58,696.00)(0.499,0.595){123}{\rule{0.120pt}{0.576pt}}
\multiput(522.17,696.00)(63.000,73.804){2}{\rule{0.400pt}{0.288pt}}
\multiput(586.00,771.58)(1.323,0.496){45}{\rule{1.150pt}{0.120pt}}
\multiput(586.00,770.17)(60.613,24.000){2}{\rule{0.575pt}{0.400pt}}
\multiput(649.00,793.92)(0.543,-0.499){113}{\rule{0.534pt}{0.120pt}}
\multiput(649.00,794.17)(61.891,-58.000){2}{\rule{0.267pt}{0.400pt}}
\multiput(712.58,734.40)(0.499,-0.659){123}{\rule{0.120pt}{0.627pt}}
\multiput(711.17,735.70)(63.000,-81.699){2}{\rule{0.400pt}{0.313pt}}
\multiput(775.00,652.92)(0.618,-0.498){99}{\rule{0.594pt}{0.120pt}}
\multiput(775.00,653.17)(61.767,-51.000){2}{\rule{0.297pt}{0.400pt}}
\multiput(838.58,599.00)(0.499,-1.082){123}{\rule{0.120pt}{0.963pt}}
\multiput(837.17,601.00)(63.000,-134.000){2}{\rule{0.400pt}{0.482pt}}
\multiput(901.00,465.92)(0.552,-0.499){111}{\rule{0.542pt}{0.120pt}}
\multiput(901.00,466.17)(61.875,-57.000){2}{\rule{0.271pt}{0.400pt}}
\multiput(964.58,407.37)(0.499,-0.667){123}{\rule{0.120pt}{0.633pt}}
\multiput(963.17,408.69)(63.000,-82.685){2}{\rule{0.400pt}{0.317pt}}
\multiput(1027.00,324.93)(4.764,-0.485){11}{\rule{3.700pt}{0.117pt}}
\multiput(1027.00,325.17)(55.320,-7.000){2}{\rule{1.850pt}{0.400pt}}
\multiput(1090.00,317.92)(2.693,-0.492){21}{\rule{2.200pt}{0.119pt}}
\multiput(1090.00,318.17)(58.434,-12.000){2}{\rule{1.100pt}{0.400pt}}
\multiput(1153.00,305.92)(1.092,-0.497){55}{\rule{0.969pt}{0.120pt}}
\multiput(1153.00,306.17)(60.989,-29.000){2}{\rule{0.484pt}{0.400pt}}
\multiput(1216.00,276.92)(0.989,-0.497){61}{\rule{0.887pt}{0.120pt}}
\multiput(1216.00,277.17)(61.158,-32.000){2}{\rule{0.444pt}{0.400pt}}
\multiput(1279.00,244.92)(1.055,-0.497){57}{\rule{0.940pt}{0.120pt}}
\multiput(1279.00,245.17)(61.049,-30.000){2}{\rule{0.470pt}{0.400pt}}
\multiput(1342.00,214.92)(1.446,-0.496){41}{\rule{1.245pt}{0.120pt}}
\multiput(1342.00,215.17)(60.415,-22.000){2}{\rule{0.623pt}{0.400pt}}
\end{picture}

\end{figure}

\end{document}